\documentstyle[sprocl,epsfig]{article}

\def\tmt{\times 10^{-2}}
\def\tmth{\times 10^{-3}}
\def\tmf{\times 10^{-4}}
\def\tmfv{\times 10^{-5}}
\newcommand{\beq}{\begin{equation}}
\newcommand{\eeq}{\end{equation}}
\newcommand{\bea}{\begin{eqnarray}}
\newcommand{\eea}{\end{eqnarray}}
\newcommand{\barr}{\begin{array}}
\newcommand{\earr}{\end{array}}
\newcommand{\bc}{\begin{center}}
\newcommand{\ec}{\end{center}}
\newcommand{\btab}{\begin{tabular}}
\newcommand{\etab}{\end{tabular}}
\newcommand{\gv}{\mbox{GeV}}
\newcommand{\tv}{\mbox{TeV}}

\newcommand{\nn}{\nonumber}
\newcommand{\ra}{\rightarrow}

\newcommand{\dro}{\Delta\rho}
\newcommand{\drqcd}{\delta\!\rho_{\rm QCD}}
\newcommand{\roro}{\rho^{(2)}}

\newcommand{\al}{\alpha}

\newcommand{\G}{\Gamma}
\newcommand{\Gmu}{G_{\mu}}

\newcommand{\ganu}{\gamma_{\nu}}

\newcommand{\gafi}{\gamma_5}

\newcommand{\noi}{\noindent}

\newcommand{\sm}{Standard Model }

\newcommand{\dal}{\Delta\alpha}

\newcommand{\mz}{M_Z^2}
\newcommand{\mw}{M_W^2}

\newcommand{\Dr}{\Delta r}

\newcommand{\alr}{A_{\rm LR}}
\newcommand{\afb}{A_{\rm FB}}

\newcommand{\ass}{asymmetries }

\newcommand{\prd}{{\it Phys.\ Rev.\ }}
\newcommand{\zp}{{\it Z.\ Phys.\ }}
\newcommand{\plb}{{\it Phys.\ Lett.\ }}
\newcommand{\pl}{{\it Phys.\ Lett.\ }}
\newcommand{\prl}{{\it Phys.\ Rev.\ Lett.\ }}
\newcommand{\np}{{\it Nucl.\ Phys.\ }}

\newcommand{\ms}{\overline{MS}}

\begin{document}

\thispagestyle{empty}
\setcounter{page}{0}
\def\thefootnote{\fnsymbol{footnote}}

\begin{flushright}
KA--TP--22--1997\\
\end{flushright}

\vspace{1cm}

\begin{center}

{\large\sc {\bf REVIEW: STATUS OF THE STANDARD MODEL}}
\footnote{talk given at the International Workshop on
          Quantum Effects in the MSSM, Barcelona,
          9 - 13  September 1997}

\vspace{1cm}

{\sc W. Hollik
}

\vspace*{1cm}

     Institut f\"ur Theoretische Physik \\ Universit\"at Karlsruhe \\
     D-76128 Karlsruhe, Germany

\vspace*{0.4cm}


\end{center}

\vspace*{1cm}

\begin{abstract}
This talk summarizes topical theoretical work for tests of the electroweak
theory and reviews
the status of the electroweak Standard Model 
in view of the recent precision  data
reported at the 1997
summer conferences.
\end{abstract}

\def\thefootnote{\arabic{footnote}}
\setcounter{footnote}{0}

\newpage

\title{REVIEW: STATUS OF THE STANDARD MODEL}

\author{ W. HOLLIK }

\address{ Institut f\"ur Theoretische Physik, Universit\"at Karlsruhe \\  
          D-76128 Karlsruhe, Germany \\
          E-mail: Wolfgang.Hollik@physik.uni-karlsruhe.de}


\maketitle\abstracts{
This talk summarizes topical theoretical work for tests of the electroweak
theory and reviews 
the status of the electroweak Standard Model 
in view of the recent precision  data
reported at the 1997
summer conferences.}

\section{Introduction}
Impressive experimental results have been obtained for 
the $Z$ boson parameters \cite{moriond}, the $W$ mass
\cite{moriond,wmass}, and the 
top quark mass \cite{top} with
$m_t = 175.6 \pm 5.5$ GeV.
The search for the Higgs boson, the only empirically missing entry
of the Standard Model, will therefore remain one of the main tasks
for testing the electroweak theory, together with tests at the
quantum level with the help  of the precision observables.

For probing the virtual effects  of the Standard Model, also
a sizeable amount of theoretical work has
contributed over the last few years to a steadily rising
improvement of the Standard Model predictions
(for a review see ref.\ \cite{yb95}). The availability of both 
highly accurate measurements and theoretical predictions, at the level
of nearly 0.1\% precision,
 provides
tests of 
the quantum structure of the \sm thereby
probing its empirically yet untested sector, and simultaneously accesses
alternative scenarios like the minimal supersymmetric extension of  
of the Standard Model, which is the issue of this workshop.

\section{Status of precision calculations}
\subsection{Radiative corrections in the Standard Model}

The possibility of performing precision tests is based
on the formulation of the \sm as a renormalizable quantum field
theory preserving its predictive power beyond tree level
calculations. With the experimental accuracy 
being sensitive to the loop
induced quantum effects, also the Higgs sector of the \sm
is probed. The higher order terms
induce the sensitivity of electroweak observables
to the top and Higgs mass $m_t, M_H$
and to the strong coupling constant $\al_s$.
 
\smallskip \noi 
Before one can make predictions from the theory,
a set of independent parameters has to be taken from experiment.
For practical calculations the physical input quantities
$ \al, \; \Gmu,\; M_Z,\; m_f,\; M_H, \; \al_s $
are commonly used    
for fixing the free parameters of the Standard Model.
 Differences between various schemes are formally
of higher order than the one under consideration.
 The study of the
scheme dependence of the perturbative results, after improvement by
resumming the leading terms, allows us to estimate the missing
higher order contributions.
 
\smallskip
Two sizeable effects in the electroweak loops deserve a special
discussion:
\begin{itemize}
\item
The light fermionic content of the subtracted photon vacuum polarization
corresponds to a QED induced shift
in the electromagnetic fine structure constant. The evaluation of the
light quark content
 \cite{eidelman}
 yield the result
\beq  (\dal)_{had} = 0.0280 \pm 0.0007\, . \eeq
Other determinations \cite{swartz}
agree within one standard deviation. Together with the leptonic
content, $\dal$ can
be resummed resulting in an effective fine structure
constant at the $Z$ mass scale:
\beq
   \al(\mz) \, =\, \frac{\al}{1-\dal}\,=\,
   \frac{1}{128.89\pm 0.09} \, .
\eeq
 \item
The electroweak mixing angle is related to the vector boson
masses  by
\bea
  \sin^2\theta  = 
   1-\frac{\mw}{\mz} + \frac{\mw}{\mz} \dro\, +  \cdots
\eea
where
the main contribution to the $\rho$-parameter 
 is from the  $(t,b)$ doublet \cite{rho},
at the present level calculated to
 \beq
 \dro= 3 x_t \cdot [ 1+ x_t \,  \roro+ \drqcd ]
\eeq
with
\beq
 x_t =
 \frac{\Gmu m_t^2}{8\pi^2\sqrt{2}} \, .
\eeq
 The electroweak 2-loop
 part \cite{bij,barbieri} is described by the
function $\roro(M_H/m_t)$.
$\drqcd$ is the QCD correction
to the leading $\Gmu m_t^2$ term
 \cite{djouadi,tarasov}
$$
    \drqcd = - 2.86 a_s  - 14.6 a_s^2, \;\;\;\;
 a_s = \frac{\al_s(m_t)}{\pi} \, .
$$

\end{itemize}
\subsection{The vector boson mass correlation}
The correlation between
the masses $M_W,M_Z$ of the vector bosons,          in terms
of the Fermi constant $\Gmu$, is in 1-loop order given by
 \cite{sirmar}:
\beq
\label{mw}
\frac{\Gmu}{\sqrt{2}}   =
            \frac{\pi\al}{ M_W^2
             \left(1-\frac{\mw}{\mz} \right) }\, 
           [ 1+ \Dr(\al,M_W,M_Z,M_H,m_t) ] \, .
\eeq
 
\medskip \noi
The appearance of large terms in $\Dr$ requires the consideration
of higher than 1-loop effects.
At present, the following  
higher order contributions are available:
\begin{itemize}
\item
The leading log resummation \cite{marciano} of $\dal$: \\
$  1+\dal\, \ra \, (1-\dal)^{-1}$
\item
The incorporation of
non-leading higher order terms
containing mass singularities of the type $\al^2\log(M_Z/m_f)$
from the light fermions \cite{nonleading}.
\item
The resummation of the leading $m_t^2$ contribution \cite{chj}
in terms of $\dro$ in Eq.\ (4).
 Moreover, the complete
 $O(\al\al_s)$ corrections to the self energies
 are available \cite{qcd,dispersion1},
 and part of the $O(\al\al_s^2)$ terms \cite{steinhauser}.

\item
The non-leading $\Gmu^2m_t^2 M_Z^2$ contribution 
of the electroweak 2-loop order \cite{padova,padova1}.
Meanwhile also the Higgs-dependence of the non-leading
$m_t$-terms has been calculated at two-loop order  
\cite{bauberger}.
\end{itemize}

\subsection{$Z$ boson observables}
With $M_Z$ as a precise input parameter, 
the predictions for the partial widths
as well as for the asymmetries
can conveniently be calculated in terms of effective neutral
current coupling constants for the various fermions:
\bea
\label{nccoup}
   J_{\nu}^{\rm NC}      &  =  &  
  g_V^f \,\ganu -  g_A^f \,\ganu\gafi     \\ 
  & = &   \left( \rho_f \right)^{1/2}
\left( (I_3^f-2Q_fs_f^2)\ganu-I_3^f\ganu\gafi \right)   \nn
\eea
with form factors 
$\rho_f$ 
for the overall normalization and the
effective mixing angles $s_f^2 \equiv \sin^2\theta_f$.

\smallskip
The effective mixing angles are of particular interest since
they determine the on-resonance asymmetries via the combinations
   \beq
    A_f = \frac{2g_V^f g_A^f}{(g_V^f)^2+(g_A^f)^2}  
\eeq
in the following way:
\beq
\label{afb}
\alr = A_e, \quad  \afb^f = \frac{3}{4}\, A_e A_f \, .
\eeq
Measurements of the \ass hence are sensitive to
the ratios
\beq
  g_V^f/g_A^f = 1 - 2 Q_f s_f^2
\eeq
or to the effective mixing angles, respectively.

\smallskip
The total
$Z$ width $\Gamma_Z$ can be calculated
essentially as the sum over the fermionic partial decay widths.
Expressed in terms of the effective coupling constants they
read up to 2nd order in the fermion masses:
\bea
\Gamma_f
  & = & \G_0
 \, \left(
     (g_V^f)^2  +
     (g_A^f)^2 (1-\frac{6m_f^2}{\mz} )
                           \right)        \nn \\
 &  & \cdot   (1+ Q_f^2\, \frac{3\al}{4\pi} ) 
          + \Delta\G^f_{QCD} \nn
\eea
with
$ \left[ N_C^f = 1
 \mbox{ (leptons)}, \;\; = 3 \mbox{ (quarks)} \right] $ 
\[
\G_0 \, =\,
  N_C^f\,\frac{\sqrt{2}\Gmu M_Z^3}{12\pi},
\]
and the QCD corrections  $ \Delta\G^f_{QCD} $
 for quark final states
 \cite{qcdq}.
The recently obtained non-factorizable part of the 2-loop 
$O(\al\al_s)$ QCD corrections \cite{czarnecki} 
yields an extra negative contribution of 
 -0.59(3) MeV for the total hadronic $Z$ width.

\subsection{Accuracy of the Standard Model predictions}
 For a discussion of the theoretical reliability
of the \sm predictions one has to consider the various sources
contributing to their
uncertainties:

The experimental error of the hadronic contribution
to $\al(\mz)$, Eq.\ (2), leads to
$\delta M_W = 13$ MeV in the $W$ mass prediction, and
$\delta\sin^2\theta = 0.00023$ common to all of the mixing
angles, which matches with the experimental precision.

The uncertainties from the QCD contributions
can essentially be traced back to
those in the top quark loops for the $\rho$-parameter.
They  can be combined into the following errors
\cite{kniehl95}:
$$
 \delta(\dro) \simeq 1.5\cdot 10^{-4},   \;
 \delta s^2_{\ell} \simeq 0.0001 \, ,
$$
which may be considered as conservative estimates. Less conservative
estimates with smaller errors are given in \cite{padova1}.
 
The size of unknown higher order contributions can be estimated
by different treatments of non-leading terms
of higher order in the implementation of radiative corrections in
electroweak observables (`options')
and by investigations of the scheme dependence.
Explicit comparisons between the results of 5 different computer codes  
based on  on-shell and $\ms$ calculations
for the $Z$ resonance observables are documented in the ``Electroweak
Working Group Report'' \cite{ewgr} in ref.\ \cite{yb95}.
Table 1  shows the uncertainty in a selected set of
precision observables.
The recently calculated 
non-leading 2-loop corrections
$\sim \Gmu^2m_t^2 M_Z^2$  \cite{padova}
for $\Delta r$ and $s_{\ell}^2$  (not included in Table 1)
reduce the uncertainty in $M_W$ and $s^2_{\ell}$ considerably,
by at least a factor 0.5.

\begin{table}[htbp] 
\caption[]
{Largest half-differences among central values $(\Delta_c)$ and among
maximal and minimal predictions $(\Delta_g)$ for $m_t = 175\,\gv$,
$60\,\gv < M_H < 1\,\tv$, $\al_s(\mz) = 0.125$
(from ref.\ \cite{ewgr}) }
\vspace{0.4cm}
\centering
\begin{tabular}{@{} c c c}
\hline 
Observable $O$ & $\Delta_c O$  & $\Delta_g O$ \\
\hline
 & & \\
$M_W\,$(GeV)          & $4.5\tmth$ & $1.6\tmt$\\
$\G_e\,$(MeV)          & $1.3\tmt$ & $3.1\tmt$\\
$\G_Z\,$(MeV)          & $0.2$     & $1.4$\\
$ s^2_e$             & $5.5\tmfv$ & $1.4\tmf$\\
$ s^2_b$             & $5.0\tmfv$ & $1.5\tmf$\\
$R_{had}$                 & $4.0\tmth$& $9.0\tmth$\\
$R_b$                 & $6.5\tmfv$ & $1.7\tmf$ \\
$R_c$                 & $2.0\tmfv$& $4.5\tmfv$ \\
$\sigma^{had}_0\,$(nb)    & $7.0\tmth$ & $8.5\tmth$\\
$\afb^l$             & $9.3\tmfv$ & $2.2\tmf$\\
$\afb^b$             & $3.0\tmf$ & $7.4\tmf$ \\
$\afb^c$             & $2.3\tmf$ & $5.7\tmf$ \\
$\alr$                & $4.2\tmf$ & $8.7\tmf$\\
 & & \\
\hline 
\end{tabular}
 
\end{table}

\section{Standard Model and precision data}
In this section we put together the \sm predictions for
the discussed set of precision observables for comparison with the most 
recent experimental data \cite{moriond,sld}.
The values for the various forward-backward
asymmetries are for the pure resonance terms (\ref{afb})
only. The small 
photon and interference contributions 
are subtracted from the data,
as well as the QED corrections.
In Table \ref{zobs}
the \sm predictions for $Z$ pole observables and the $W$ mass  are
put together for a light and a heavy Higgs particle with $m_t=175$ GeV.
 The last column is the variation of the prediction according to
$\Delta m_t = \pm 6$ GeV. The input value 
for $\al_s$ is chosen as $\al_s = 0.118$ \cite{schmelling}.
Not included are the uncertainties from
$\delta\al_s=0.003$, which amount to 1.6 MeV for the hadronic $Z$ width,
0.038 nb for the hadronic peak cross section, and 0.019 for $R_{had}$.
The other observables are insensitive to small variations of $\al_s$.
The experimental results on the $Z$ observables are from 
LEP and SLD ($A_b, A_c$ and $s_e^2$ from $\alr$).
The leptonic mixing angle determined via $\alr$ by SLD and the 
$s^2_{\ell}$ average 
from LEP (assuming lepton universality)
differ by about 3 standard deviations:
\bea
   s^2_e(\alr) & = & 0.23055 \pm 0.00041  \nn \\
   s^2_{\ell} ({\rm LEP}) & = & 0.23196 \pm 0.00028 \, . \nn
\eea
Table \ref{zobs} contains the combined LEP/SLD value 
for $s_e^2$ as given by the LEP
Electroweak Working Group \cite{moriond}.
Alternatively, in view of the obvious discrepancy,
it has been proposed \cite{altarelli97}
to enlarge the error by the factor 
$\sqrt{\chi^2/{\rm d.o.f}} = \sqrt{12.5/6}
 \simeq 1.4$ to $\pm 0.0032$ for a more 
conservative error estimate following the Particle Data Group.
 $\rho_{\ell}$ and $s^2_{\ell}$ in Table \ref{zobs} 
are the leptonic
neutral current couplings in eq.~(\ref{nccoup}), 
derived from partial widths and
asymmetries  under the assumption of lepton universality.
The table illustrates the sensitivity of the various quantities 
to the Higgs mass.
The effective mixing angle turns out to be
the most sensitive  observable, where both the experimental error and the
uncertainty from $m_t$ are small compared to the variation with $M_H$.
Since a light Higgs boson  corresponds to
a low value of $s^2_{\ell}$,
the strongest upper bound on $M_H$ is from $\alr$ at the SLC \cite{sld},
 whereas 
LEP data alone allow to accommodate also a relatively heavy Higgs 
(see Figure \ref{fig3}).
Further constraints on $M_H$ are to be expected in the future from
more precise $M_W$ measurements at LEP 2 \cite{ww}
and the upgraded Tevatron.

\begin{table*}[t]
\vspace*{0.5cm} 
            \caption{\label{zobs}Precision observables: 
              experimental results 
             {\protect\cite{moriond}}
             and \sm         
             predictions. }
\vspace{0.4cm}
\begin{center}
 \btab{@{} l  l  r  r  r  }
\hline 
observable & exp.  & $M_H= 65$ GeV & 1 TeV &
 $ \Delta m_t $ \\
\hline
 & & & & \\
$M_Z$ (GeV) & $91.1867\pm0.0020$ &  input & input &    \\
$\Gamma_Z$ (GeV) & $2.4948\pm 0.0025$ & 2.4974 & 2.4881 & $\pm 0.0015$ \\
$\sigma_0^{had}$ (nb) & $41.486\pm 0.053$ & 41.476 & 41.483 & $\pm 0.003$  \\
 $ R_{had}$ & $20.775\pm 0.027 $ & 20.753 & 20.725 
                                                     & $\pm 0.002$ \\
$R_b$  & $0.2170\pm 0.0009$ & 0.2156 & 0.2157 & $\pm 0.0002$ \\
$ R_c$  & $0.1734\pm0.0048$ & 0.1724 & 0.1723 & $\pm 0.0001$ \\
$\afb^{\ell}$ & $0.0171 \pm 0.0010$ & 0.0170 & 0.0144 & $\pm 0.0003$ \\
$\afb^b$ & $0.0983 \pm 0.0024$ &  0.1056 & 0.0970 & $\pm 0.0010$ \\
$\afb^c$ & $0.0739 \pm 0.0048$ &  0.0756 & 0.0689 & $\pm 0.0008$ \\
$A_b$            & $0.900\pm 0.050$  & 0.9340 & 0.9350 &  $\pm 0.0001$ \\
$A_c$            & $0.650\pm 0.058$  & 0.6696 & 0.6638 &  $\pm 0.0006$ \\
$\rho_{\ell}$ & $1.0041\pm 0.0012$ & 1.0056 & 1.0036 & $\pm 0.0006$ \\
$s^2_{\ell}$  & $0.23152\pm 0.00023$ & 0.23114 & 0.23264 & $\pm 0.0002$ \\
$M_W$ (GeV) & $80.43 \pm 0.08$ & 80.417 & 80.219 & $\pm 0.038$  \\
  & & & & \\
\hline
\etab
 \vspace{0.5cm}
\end{center}
\clearpage
\end{table*}

\smallskip 
Note that
the experimental value for $\rho_{\ell}$ points out  the presence of
genuine electroweak corrections by 3.4 standard deviations 
($\rho_{\ell}=1$ at tree level).
The 
deviation from the \sm prediction in the quantity $R_b$ has been reduced
to about one standard deviation by now.
Other small deviations 
 are observed in the asymmetries: the purely leptonic $\afb$ is slightly
higher than the \sm predictions, and $\afb$ for $b$ quarks is 
lower.
Whereas the leptonic $\afb$ favors a very light Higgs boson, 
the $b$ quark asymmetry needs a heavy Higgs. The measured asymmetries deviate
from the best fit  values by 2 and 2.4  standard deviations in the 
worst cases of $\afb^b$ and $\alr$ (Table \ref{zobs1}).

\smallskip 
The $W$ mass prediction in Table \ref{zobs}
is obtained  from Eq.~(\ref{mw}) 
(including the higher order terms)
from
 $M_Z,\Gmu,\al$ and  $M_H,m_t$.
The present experimental value for the $W$ mass
from the combined UA2,  CDF and D0 results \cite{wmass} is
\beq
\label{mwpp}
  M_W^{\rm exp}\, = \, 80.41 \pm 0.09 \, \gv \ , 
\eeq
and from LEP 2 \cite{moriond}:
\beq
\label{mwlep}
  M_W^{\rm exp}\, = \, 80.48 \pm 0.14 \, \gv \ , 
\eeq
yielding the average given in Table \ref{zobs}.

\smallskip
The quantity $s_W^2$ resp.\  the ratio $M_W/M_Z$
can indirectly be measured in deep-inelastic
neutrino-nucleon scattering.
The present world average on $s_W^2$ from the experiments
CCFR, CDHS and CHARM \cite{neutrino} 
\beq
\label{sw}
s_W^2 = 1 - M_W^2/M_Z^2 = 0.2236 \pm 0.0041   
\eeq
corresponds to $M_W = 80.35 \pm 0.21$ GeV and hence 
is fully consistent with the direct vector boson mass measurements
and with the standard theory. 

\smallskip \noi
The indirect determination of the       
$W$ mass  from the global fit to the LEP1/SLD data \cite{moriond}
\[ M_W = 80.329\pm 0.041 \, \gv \, , \]
is slightly lower than the experimental world average, but still
in agreement with the direct measurement.

\bigskip \noi
{\it Standard Model global fits:}

\begin{table*}[t]
\vspace*{0.5cm} 
            \caption{\label{zobs1}Precision observables: 
              experimental results 
             and \sm best fit values
             {\protect\cite{moriond}}. }
\vspace{0.4cm}
\begin{center}
 \btab{@{} l  l  l  r  }
\hline 
observable & exp.  &  best fit  & pull  \\
\hline
 & & &  \\
$M_Z$ (GeV) & $91.1867\pm0.0020$ & 91.1866 &  0.0    \\
$\Gamma_Z$ (GeV) & $2.4948\pm 0.0025$ & 2.4966 & -0.7 \\
$\sigma_0^{had}$ (nb) & $41.486\pm 0.053$ & 41.467 & 0.4 \\
 $ R_{had}$ & $20.775\pm 0.027 $ & 20.756 & 0.7 \\
$R_b$  & $0.2170\pm 0.0009$ & 0.2158 & 1.4 \\
$ R_c$  & $0.1734\pm0.0048$ & 0.1723 & -0.1 \\
$\afb^{\ell}$ & $0.0171 \pm 0.0010$ & 0.0162 & 0.9 \\
$\afb^b$ & $0.0983 \pm 0.0024$ &  0.1031 & -2.0 \\
$\afb^c$ & $0.0739 \pm 0.0048$ &  0.0736 & 0.0 \\
$A_b$            & $0.900\pm 0.050$  & 0.935 & -0.7 \\ 
$A_c$            & $0.650\pm 0.058$  & 0.668 & -0.3 \\
$\rho_{\ell}$ & $1.0041\pm 0.0012$ & 1.0054 & -1.0 \\
$s^2_{\ell}$ (LEP)  & $0.23196\pm 0.00028$ & 0.23152 & 1.6 \\
$s^2_{\ell}$ (SLD)  & $0.23055\pm 0.00041$ & 0.23152 & -2.4 \\
$M_W$ (GeV) & $80.43 \pm 0.08$ & 80.375 & 0.7 \\
  & & & \\
\hline
\etab
 \vspace{0.5cm}
\end{center}
\clearpage
\end{table*}

\begin{figure}[htb]
\centerline{
\epsfig{figure=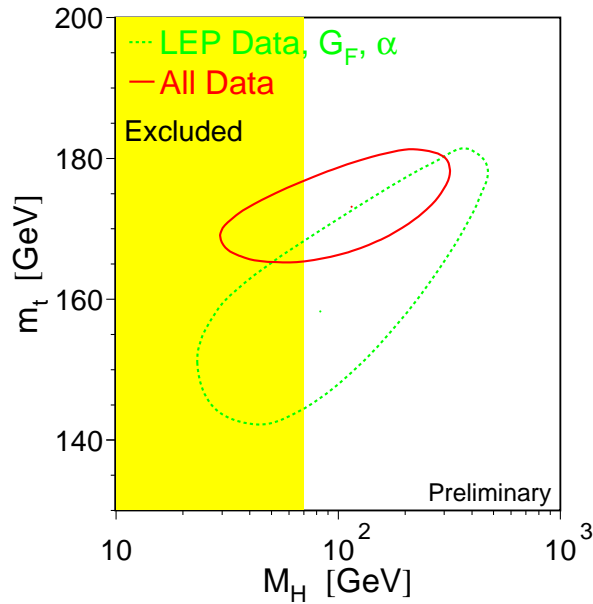,height=8cm}}
\caption{The 68\% confidence level contours in the top and Higgs masses
         for the  fits to the LEP data only (dashed) and to all data
         (solid) including $m_t$ measurements
         (from ref.\ \protect \cite{moriond})}
\label{mtmh}
\end{figure}

\begin{figure}[htb]
\vspace{-1cm}
\centerline{
\epsfig{figure=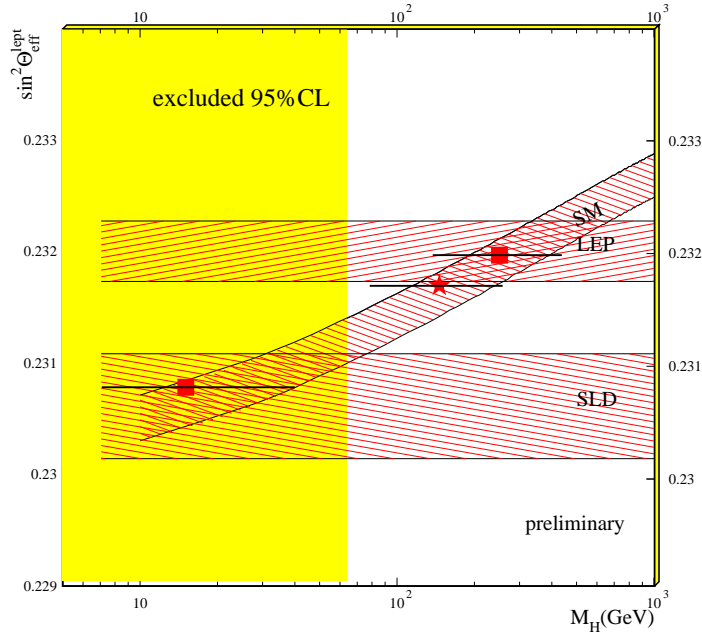,height=12cm}}
\vspace{-1.5cm}
\caption{Dependence of the leptonic mixing angle on the Higgs mass.
         The theoretical predictions correspond to
         $m_t=175\pm 6$ GeV. The SLD 
         and LEP 
         measurements are separately
         shown. The star is the result of a combined fit to LEP and SLD
         data, the squares are for separate fits 
         (from ref.\ {\protect\cite{deboer}}) } 
\label{fig3}
\end{figure}

\smallskip  \noi
In the meantime the data have reached an accuracy such that
global fits with 
respect to both $m_t$ and $M_H$ as free parameters have become 
available \cite{moriond,haidt,higgsfits,deboer} (Figure \ref{mtmh}).
The results of ref.~\cite{moriond} based on the
most recent  LEP (still preliminary) and SLD data
are
\[ m_t = 157^{+10}_{-9}\, \gv, \quad 
   M_H = 41^{+64}_{-21} \, \gv \, .
\]
Without the
Tevatron constraint on the top mass, the favored range for $m_t$ is
a bit lower than the direct measurement. The reason for this
behaviour is the strong impact on the
upper limit of $m_t$ from the quantity $R_b$.

Treating the top mass as an additional experimental data point,
the global fit to all electroweak results from
LEP, SLD, $M_W$,  $\nu N$ and $m_t$ 
yields  the following results \cite{moriond} 
for $m_t$ and $\al_s$
\[ m_t = 173.1 \pm 5.4 \, \gv, \quad 
   \al_s = 0.120 \pm 0.003 \]
and for the Higgs mass
\[ M_H = 115^{+116}_{-66} \, \gv  \]
with an overall  $\chi^2 = 17/15$.
The value obtained for
 $\al_s$ is in very good agreement with the world average
\cite{altarelli97,schmelling}.
The input from $\alr$
is decisive for a restrictive upper bound for $M_H$. 
Without $\alr$,  the 95\% C.L  
upper bound is shifted upwards by more
than 200 GeV \cite{deboer}.

\smallskip 
The numbers given above 
 do not yet include the theoretical uncertainties of the
\sm  predictions. The LEP Electroweak Working Group
 \cite{moriond}
has performed a study of the influence of the various `options'
discussed in section 2.4 on the bounds for the Higgs mass with the result
that the 95\% C.L. one-sided upper bound is shifted by nearly 
100 GeV to higher values, yielding
\[ M_H < 420 \, \gv \, (95\% \, {\rm C.L.}) \, . \]
It has to be kept in mind, however, that this error estimate is based on the
uncertainties as given in Table 1. Since the recent improvement in the 
theo\-retical prediction \cite{padova}
is going to reduce the theo\-retical uncertainty
one may expect also a significant
smaller theoretical error on the Higgs mass bounds once the 2-loop terms
$\sim \Gmu^2 m_t^2 \mz$ are implemented in the codes used for the fits.
At the present stage the analysis is done without the new terms.

\smallskip 
The error from  the hadronic vacuum polarization is 
incorporated in the fit and is thus part of the result on the Higgs mass
bound. The uncertainty induced from $\dal$ is 
quite remarkable at the present stage (see for example the discussion
in \cite{haidt}).

\medskip
There are also  theoretical constraints on the Higgs mass
from vacuum stability and absence of a Landau pole \cite{lindner},
and from lattice calculations \cite{lattice}. Recent calculations
of the decay width for $H\ra W^+W^-,ZZ$  in the large $M_H$ limit
in 2-loop order \cite{ghinculov} have  shown that the 2-loop
contribution exceeds the 1-loop term in size (same sign) for
 $M_H > 930$ GeV. The requirement of applicability of
perturbation theory therefore puts a stringent upper limit on the
Higgs mass \cite{riesselmann}.
The indirect Higgs mass bounds obtained from the
precision analysis show, however, that the Higgs boson is well below
the mass range where the Higgs sector becomes non-perturbative.

\section{Conclusions}
The experimental data for tests of the \sm
have achieved an impressive accuracy.
In the meantime, many 
theoretical contributions have become available to improve and 
stabilize the \sm predictions. To reach, however, a theoretical
accuracy at the level of 0.1\% or below, new experimental data on
$\dal$ and more complete electroweak 2-loop calculations are required.

The agreement of the 
electroweak precision data with
the \sm predictions is remarkably good. The quality of a global fit,
converted into a probability for the Standard Model of 
about  31\%, 
has even improved over the last two years by both experimental and
theoretical efforts.
The description of the current data by the minimal model 
is thus extraordinarily successful. The few observed deviations can 
be understood as fluctuations which appear as statistically normal.
A further  check of the consistency of the theory is the description 
of the entire  set of
precision observables with a light Higgs boson which lies  
well below the non-perturbative
regime,
confirming the perturbative character of the electroweak
\sm which might be extrapolated 
up to high energies
compatible with the Planck scale.
A light Higgs boson, however, can also 
be naturally attributed to a supersymmetric
extension of the \sm like the MSSM,
which provides a competitive
description of the current precision data with a similar quality as
the Standard Model \cite{hollik}.

\section*{Acknowledgements}
I want to thank Joan Sol\`a and the Organizing Committee for the 
invitation to this Workshop and 
for the kind hospitality at the
Universitat Autonoma de Barcelona.

\end{document}